\documentclass[manuscript,noblind]{geophysics}
\usepackage{amsmath}
\usepackage[colorlinks,
linkcolor=blue,
urlcolor = blue,
citecolor=blue
]{hyperref}

\begin{document}

\title{Fast Diffusion Model For Seismic Data Noise Attenuation}

\renewcommand{\thefootnote}{\fnsymbol{footnote}} 

\ms{Submitted to Geophysics} 

\address{
\footnotemark[1]The Key Laboratory of Earth Exploration \& Information Techniques of Ministry Education \\
Chengdu University of Technology \\
Chengdu, Sichuan, China \\
\footnotemark[2]School of Geophysics \\
Chengdu University of Technology \\ 
Chengdu, Sichuan, China \\}

\author{Junheng Peng\footnotemark[1]\footnotemark[2], Yong Li\footnotemark[1]\footnotemark[2], Yingtian Liu\footnotemark[1]\footnotemark[2], Zhangquan Liao\footnotemark[1]\footnotemark[2]}

\footer{Example}
\lefthead{Peng et al.}
\righthead{\emph{Fast Diffusion Model}}

\maketitle

\begin{abstract}
  Noise is one of the primary sources of interference in seismic exploration. Many authors have proposed various methods to remove noise from seismic data; however, in the face of strong noise conditions, satisfactory results are often not achievable. In recent years, methods based on diffusion models have been applied to the task of strong noise processing in seismic data. However, due to iterative computations, the computational efficiency of diffusion-based methods is much lower than conventional methods. To address this issue, we propose using an improved Bayesian equation for iterations, removing the stochastic terms from the computation. Additionally, we proposed a new normalization method adapted to the diffusion model. Through various improvements, on synthetic datasets and field datasets, our proposed method achieves significantly better noise attenuation effects compared to the benchmark methods, while also achieving a several-fold increase in computational speed. We employ transfer learning to demonstrate the robustness of our proposed method on open-source synthetic seismic data and validate on open-source field data sets. Finally, we open-sourced the code to promote the development of high-precision and efficient seismic exploration work.

\end{abstract}

\section{Introduction}

In seismic exploration, the seismic data are adversely affected by random noise originating from various sources due to the influence of different environments. This interference significantly hampers the interpretation of seismic data, especially in high-precision exploration work \cite[]{1}. Conventional approaches to addressing random noise can be broadly categorized into two main types: conventional computational methods and methods based on deep learning (DL). Among them, conventional computational methods can be further divided into three categories: methods based on filtering, methods based on sparse transform, and methods based on rank reduction.

Methods based on filtering leverage the distinct characteristics of seismic signals and noise within the frequency (F-X) domain, operating under the premise that seismic events exhibit a certain degree of coherence or regularity that can be distinguished from the more erratic behavior of random noise \cite[]{1} \cite[]{2}. Building on this foundational theory, numerous authors have proposed improvements to enhance the effectiveness of filtering methods. These enhancements aim to address the limitations of traditional filtering approaches and adapt to the complexities of seismic data and the diverse nature of noise encountered in field scenarios. \cite{3} introduced the empirical mode decomposition (EMD) method to improve the performance of predictive filtering in the F-X domain; \cite{4} proposed a data-dependent Fourier filter based on image segmentation for seismic noise attenuation. Methods based on sparse transform share a similar objective with filtering-based methods, aiming to convert seismic data into sparse domains for better separation of seismic data and noise. Sparse transform methods transform seismic data into a domain where the signal can be represented as sparsely as possible, meaning that the essential information or features of the signal are concentrated in a few non-zero coefficients, while the majority of the domain remains zero or near-zero. This characteristic makes it easier to distinguish between signal and noise, as noise typically does not exhibit such a sparse representation \cite[]{5}. Common sparse transform methods include the Fourier transform \cite[]{6} \cite[]{7}, seislet transform \cite[]{8} \cite[]{9}, wavelet transform \cite[]{10} \cite[]{11} \cite[]{12}, and curvelet transform \cite[]{13} \cite[]{14}.

Methods based on rank reduction assume that the seismic data is a low rank structure, and the addition of random noise increases the rank of the data. Based on this assumption, random noise can be eliminated by employing low rank constraints \cite[]{15} \cite[]{16}. Many authors have proposed various improved methods based on this idea, \cite{17} have developed a new multichannel singular spectrum analysis algorithm to decompose the vector space of the Hankel matrix of the noisy signal into a signal subspace and a noise subspace for random noise attenuation; Based on this method, \cite{18} introduced a damping operator to further improve the noise removal performance. However, such methods are faced with certain challenges. Firstly, their computations are highly complex, particularly when dealing with high-density seismic data. Secondly, there is a certain degree of signal leakage that occurs during the data processing.

Furthermore, DL methods have been widely utilized in recent years with the advancement of the computer field \cite[]{19}. Among them, convolutional neural networks (CNN) such as ResNet \cite[]{20}, are the most widely used and achieved success in the fields of computer vision and natural language processing \cite[]{21}. On this basis, many authors have proposed improved CNN-based methods for seismic noise attenuation. \cite{22} proposed to use denoising CNN (DnCNN) to learn noise extraction instead of noise attenuation in blind Gaussian noisy data; \cite{23} systematically improved ResNet, transformed the target function from signal learning to noise learning and improved the training efficiency; Based on residual learning, \cite{24} introduced the concept of iterative processing, further enhancing the noise attenuation effectiveness. Besides, self-supervised DL methods are also worth exploring. In contrast to the aforementioned methods, self-supervised DL methods do not require extensive training data or excessive training, thereby expanding their applicability. Among them, denoising autoencoder (DAE) is the most widely applied method \cite[]{25} \cite[]{26}. Instead of using synthetic noise-free data or denoised results via conventional methods as training labels, \cite{27} proposed to use normalization and patch sampling to build training dataset and test dataset from raw noisy data; \cite{28} proposed to attenuate random noise based on a deep DAE (DDAE), and they conducted supervised pre-training on synthetic data and then transferred to unsupervised training on field data; \cite{29} proposed a denoising framework with twice DAE (TDAE) for random seismic noise attenuation to mitigate both signals leaking and noise remaining exist. Despite the advancements made by DL methods represented above in various aspects over conventional computational methods, they still struggle to achieve significant improvement when confronted with strong noise.

In recent years, denoising diffusion probabilistic model (DDPM) has achieved significant success in the field of image generation by transforming the task into a step-by-step denoising process \cite[]{30} \cite[]{31}. DDPM constructs a Markov process to generate images step by step from pure noise using Bayesian equation. However, applying DDPM directly to noise attenuation tasks in seismic data is infeasible due to the unknown noise level in seismic data, which directly determines the number of steps in the reverse process. To address this issue, \cite{32} proposed using principal component analysis (PCA) to estimate the noise level in seismic data. They established a relationship between the noise level and the number of steps in the reverse process through a cubic fitting equation. In the experiments, the effectiveness of DDPM has indeed surpassed that of most conventional computational methods and DL methods. Even in cases of extreme noise, DDPM can reconstruct the majority of signals. However, DDPM still has limitations: the establishment of the Markov process necessitates step-by-step computations during the reverse process, leading to lower efficiency compared to other methods.

In this study, inspired by the denoising diffusion implicit model (DDIM)\cite[]{33}, we propose the use of an improved DDPM reverse process. In the improved reverse process, there is no longer a strict requirement for step-by-step sampling following DDPM; instead, the reverse process can be conducted by skipping steps at certain intervals. Building upon this, we further investigated the impact of different intervals on the processing results under various noise levels, as well as the differences in computational efficiency. In this experiment, we utilized an Intel i7 10750H laptop CPU with 16GB memory and a Nvidia RTX 2060 laptop GPU with 6GB memory for our computations. The experiments demonstrate that, compared to the previously proposed DDPM method for seismic data noise attenuation, the method presented in this study exhibits a several-fold improvement in computational efficiency, while also achieving enhanced processing results. Furthermore, the proposed method exhibits improvements in comparison to comparative DL method and conventional computational methods, particularly in scenarios with extremely strong noise.

\section*{Theory}

\subsection{Diffusion Model}

The theory of noise diffusion is mainly divided into two parts, the forward process and the reverse process. Building upon this, we propose the use of an improved reverse computing, which enables the acceleration of the reverse process.

\subsubsection{Forward Process}

The essence of our work is the process of inverting the Bayesian equation to remove noise, thereby reducing noise. Therefore, in the forward process, we primarily focus on modeling the noise. Let the clean seismic data be $x_0$, and the noisy seismic data be $x_{t}$. We assume that the noise from $x_0$ to $x_t$ increases gradually, and the additional noise intensity in each step set to a sequence $\beta$. Meanwhile, to minimize errors as much as possible, we set the $\beta$ values to be very small, and the length of $\beta$ sequence is sufficiently long.

So the $x_{t}$ can be represented as:
\begin{equation}
x_t=\sqrt{\alpha_t}*x_{t-1} + \sqrt{1-\alpha_t}*z_t,\quad z_t\sim N(0,1)
\end{equation}
Among them:
\begin{equation}
\alpha_t=1-\beta_t
\end{equation}
$z_t$ represents noise at $t$ step. At the same time, the $x_{t-1}$ can also be represented as:
\begin{equation}
x_{t-1}=\sqrt{\alpha_{t-1}}*x_{t-2} + \sqrt{1-\alpha_{t-1}}*z_{t-1},\quad z_{t-1}\sim N(0,1)
\end{equation}
Bringing Equation 3 into Equation 1 yields:
\begin{equation}
x_t=\sqrt{\alpha_{t-1}*\alpha_t}*x_{t-2} + 
\sqrt{(1-\alpha_{t-1})*\alpha_t}*z_{t-1} + \sqrt{1-\alpha_t}*z_t
\end{equation}
where both $z_t$ and $z_{t-1}$ follow Gaussian distribution. Because the sum of two independent Gaussian distributions is also a Gaussian distribution, we use $z^`$ to represent both of them, which can be expressed as:
\begin{equation}
\sqrt{(1-\alpha_{t-1})*\alpha_t}*z_{t-1} + \sqrt{1-\alpha_t}*z_t = \sqrt{1-\alpha_{t-1}*\alpha_t}*z^`
\end{equation}
Thus, $x_t$ can be represented as:
\begin{equation}
x_t=\sqrt{\alpha_{t-1}*\alpha_t}*x_{t-2} + \sqrt{1-\alpha_{t-1}*\alpha_t}*z^`
\end{equation}
In the same way, multiple substitutions can yield:
\begin{equation}
x_t=\sqrt{\overline{\alpha}_t}*x_{0} + \sqrt{1-\overline{\alpha}_t}*\hat{z}_t
\end{equation}
where $\overline{\alpha}_t$ represents the cumulative multiplication of $\alpha_1$ to $\alpha_t$ and $\hat{z}_t$ represents the combination of $z_1$ to $z_t$ Gaussian distributions. Therefore, throughout the entire forward process, we derived the formula from $x_0$ to $x_t$, and similarly, we can obtain formulas with a similar structure from $x_0$ to $x_{t-1}$, and so on. When dealing with field noisy data, we estimate the noise level in the data and gradually reverse the field data based on the iterative formula mentioned above.

In the following sections, we will derive the method for calculating $x_{t-1}$ using $x_t$, gradually restoring it to obtain $x_0$, which represents the clean seismic data.

\subsubsection{Reverse Process}
In the process of reverse, we reversely solve $x_{t-1}$ through the Bayesian equation, which can be expressed as:
\begin{equation}
P(x_{t-1}|x_t, x_0)=P(x_t|x_{t-1}, x_0)*\frac{P(x_{t-1},x_0)}{P(x_t,x_0)}
\end{equation}
The terms in the equation can be computed as:
\begin{equation}
P(x_t|x_{t-1}, x_0)=\sqrt{\alpha_{t}}*x_{t-1} + \sqrt{1-\alpha_{t}}*z_{t}
\end{equation}
\begin{equation}
P(x_{t-1},x_0)=\sqrt{\overline{\alpha}_{t-1}}*x_{0} + \sqrt{1-\overline{\alpha}_{t-1}}*\hat{z}_{t-1}
\end{equation}
\begin{equation}
P(x_t,x_0)=\sqrt{\overline{\alpha}_{t}}*x_{0} + \sqrt{1-\overline{\alpha}_{t}}*\hat{z}_{t}
\end{equation}

And then, we expand them using the Gaussian distribution function:
\begin{equation}
x\sim N(\mu, \sigma^2)
\end{equation}
\begin{equation}
f(x)=\frac{1}{\sqrt{2*\pi}*\sigma}*exp(-\frac{(x-\mu)^2}{2*\sigma^2})
\end{equation}

Expanding the left-hand side of the equation similarly and simplifying, we can arrive at the following conclusion:
\begin{equation}
\mu_{t-1} = \frac{1}{\sqrt{\alpha_t}}*(x_t - \frac{1-\alpha_t}{\sqrt{1-\overline{\alpha}_t}}*\hat{z_t})
\end{equation}
\begin{equation}
\sigma_{t-1}^2=\frac{(1-\overline{\alpha}_{t-1})*\beta_t}{1-\overline{\alpha}_{t}}
\end{equation}
\begin{equation}
x_{t-1}\sim N(\mu_{t-1}, \sigma_{t-1}^2)
\end{equation}

Therefore, based on the above conclusion, by continuously reverse calculating, DDPM can gradually eliminate noise in seismic data step by step. It is also this iterative computation that significantly reduces the processing efficiency of DDPM. Especially when the noise level in seismic data is high, leading to a larger corresponding $t$, implying more iterations. In the next section, we will propose improvements to address this issue.

\subsection{Fast Reverse Process}
\subsubsection{Fast Reverse Computing}

Because the reverse calculation has to be carried out one by one, the above reverse process requires a longer computing time, which limits the application of this method. To improve the computation speed of this method, we abandon optimization equation 8 and calculate the following equation 17 instead.
\begin{equation}
P(x_{s}|x_t, x_0)=P(x_t|x_{s})*\frac{P(x_{s},x_0)}{P(x_t,x_0)},\quad s<t-1\ and\ s\neq 0
\end{equation}
among them, the following two can be easily obtained:

\begin{equation}
P(x_{s},x_0)=\sqrt{\overline{\alpha}_s}*x_{0} + \sqrt{1-\overline{\alpha}_s}*\hat{z}_s
\end{equation}
\begin{equation}
P(x_{t},x_0)=\sqrt{\overline{\alpha}_t}*x_{0} + \sqrt{1-\overline{\alpha}_t}*\hat{z}_t
\end{equation}

But $P(x_t|x_{s})$ is difficult to obtain by direct calculation, so we decided to wait for the form of $P(x_{s}|x_t, x_0)$ first:

\begin{equation}
P(x_{s}|x_t, x_0)=K*x_0+M*x_t+N*e
\end{equation}

$e$ represents Gaussian distributed noise. Substituting Equation 19 yields the following formula:
\begin{equation}
P(x_{s}|x_t, x_0)=K*x_0+M*(\sqrt{\overline{\alpha}_t}*x_{0} + \sqrt{1-\overline{\alpha}_t}*\hat{z}_t)+N*e
\end{equation}

Collating the above equation and comparing Equation 18, the following conclusions can be obtained:
\begin{equation}
K+M*\sqrt{\overline{\alpha}_t}=\sqrt{\overline{\alpha}_s}
\end{equation}
\begin{equation}
\sqrt{1-\overline{\alpha}_s}*\hat{z}_s=\sqrt{(M^2*(1-\overline{\alpha}_t)+N^2)}*e
\end{equation}
where $\hat{z}_s$ and $e$ are both Gaussian random noises, and their mean values are both 0, so their coefficients are equal. Combining Equations 20, 22, and 23, the following conclusions can be obtained:
\begin{equation}
x_s=(\sqrt{\overline{\alpha}_s}-\frac{\sqrt{1-\overline{\alpha}_s-N^2}}{\sqrt{1-\overline{\alpha}_t}}*\sqrt{\overline{\alpha}_t})*x_0+(\frac{\sqrt{1-\overline{\alpha}_s-N^2}}{\sqrt{1-\overline{\alpha}_t}})*x_t+N*e
\end{equation}

By substituting Equation 19 and subtracting $x_0$, we get the equation for $x_s$. In the derivation of the above formula, there is no involvement of $N$, theoretically allowing $N$ to take on any value within a certain range. Drawing from various works in the field of image generation \cite[]{34} \cite[]{35} \cite[]{36}, the choice of $N$ impacts the diversity of generated results. However, since the objective of seismic data noise attenuation is unique compared to image generation goals, and does not require significant diversity. So in our current work, to ensure computational efficiency, we choose $N=0$. Therefore, the final formula for the inverse process can be presented as:
\begin{equation}
x_s=\sqrt{\overline{\alpha}_s}*\frac{x_t-\sqrt{1-\overline{\alpha}_s}*\hat{z_t}}{\sqrt{\overline{\alpha}_t}}+\sqrt{1-\overline{\alpha}_s}*\hat{z_t}, \quad s<t-1
\end{equation}

In summary, in fast reverse computing, the skip-step calculation of the reverse process is realized, which improves the calculation speed of the diffusion model. In general, fast reverse computation essentially involves extracting a subchain from the complete forward process Markov chain for reverse computation as shown in Figure ~\ref{fig:1}.

In theory, if the DL model we used is precise enough to accurately predict $\hat{z}_t$, the above method can directly compute $x_0$ in reverse. However, the DL model always has a certain degree of error, and there are computational errors in multiple iterations. Therefore, we will explore the impact of different subchain lengths on the processing speed and processing effectiveness to find the optimal subchain length. We conducted experiments using a validation set consisting of 100 data samples and synthesized noisy seismic data with four different levels of noise: $t=50$, $t=100$, $t=150$, and $t=200$. The relationship between signal-to-noise ratio (SNR) and the length of the subchain extracted during fast reverse computation is shown in Figure ~\ref{fig:2}.

During fast reverse computation, the minimum length of the extracted subchain is 3, including $x_0$, $x_1$, and the input $x_t$. It can be observed that when the noise level is low ($t=50$), extracting a subchain of length 3 for reverse computation yields good SNR results. However, as the subchain length increases, the number of steps required for calculation increases as well. The errors accumulated due to model predictions of $\hat{z}_t$ lead to a decrease in SNR. With the increase in seismic data noise level, extracting a subchain of length 3 does not yield optimal results; instead, a longer subchain is required. Beyond the optimal length of the subchain, there is also a decrease in SNR due to error accumulation. Based on this, we supplemented more experiments on the optimal subchain length under different noise levels. The experimental results are shown in Figure ~\ref{fig:3}.

What can be seen is that, compared to the gradual iteration of DDPM, in the fast reverse process, it is only necessary to select subchains of lengths 3-5 for the reverse process, which can also achieve good results. In addition, there was no significant difference in the results of subchains of length 4 or 5 when $t$ was around 70 to 75 and 170 to 175. Therefore, in this study, we calculate the optimal length of subchains during the reverse process using the following formula:
\begin{equation}
L=\left\{
\begin{aligned}
&  3\quad t\leq 75 \\
&  4\quad 75\textless t\leq 175  \\
&  5\quad 175\textless t\leq 200
\end{aligned}
\right.
\end{equation}
where $L$ represents the length of the optimal subchain. To ensure stable results, given the optimal subchain length $L$, we select the most dispersed subchain for the reverse process, which can be represented as:
\begin{equation}
Reverse(i)=\left\{
\begin{aligned}
&  0\quad i=1 \\
&  1+(i-2)*\lceil\frac{t-1}{L-2} \rceil \quad 1\textless i\textless L \\
&  t \quad i=L
\end{aligned}
\right.
\end{equation}

In this section, we first derived the formula for fast reverse computation, essentially extracting a subchain from the Markov chain of DDPM for the reverse process. Secondly, experiments show that subchain lengths between 3 to 5 are sufficient to remove noise from seismic data. Subchains that are too short will reduce the SNR of processing results due to insufficient accuracy in identifying $\hat{z}_t$ by the DL model, while subchains that are too long will lead to error accumulation, also reducing the SNR of processing results. Based on the experiments mentioned above, we have defined the formula for subchains to be used in the upcoming experiments.

\subsubsection{Normalization and Restore}
Due to the different numerical ranges in earthquake data, normalization is required during processing, and then the data needs to be restored after processing. Due to the unique nature of the Markov process we are modeling, traditional normalization methods are not applicable. Therefore, we propose a new normalization approach. Suppose the noisy seismic data satisfies the following conditions:
\begin{equation}
x_0=\mu_0+\sigma_0*e_0,\quad e_0\sim N(0,1)
\end{equation}
\begin{equation}
Data_{noisy}=\mu_{noisy}+\sigma_{noisy}*e_{noisy},\quad e_{noisy}\sim N(0,1)
\end{equation}

In this case, we utilize the PCA estimation method proposed \cite[]{32} in previous work to estimate $t$ for the noisy data. The $\mu_0$ and $\sigma_0$ are obtained by calculating the statistical average on the training set:
\begin{equation}
\mu_0=5.0297*10^{-4},\quad \sigma_0^2=1.0853
\end{equation}

Under the training set conditions, combining equations 7 and 28, we can derive that the noisy data corresponding to $x_t$ satisfies:
\begin{equation}
x_t=\sqrt{\overline{\alpha}_t}*\mu_0+\sqrt{\overline{\alpha}_t*\sigma_0^2+1-\overline{\alpha}_t}*\hat{e},\quad \hat{e}\sim N(0,1)
\end{equation}
where $\hat{e}$ is the standard normal distribution term resulting from the combination of $e_0$ and $\hat{z}_t$.

Therefore, our goal is to normalize the noisy seismic data to its corresponding form at $t$. Thus, the normalization formula can be represented as:
\begin{equation}
Data_{norm}=\frac{Data_{noisy}-Mean(Data_{noisy})}{Var(Data_{noisy})}*\sqrt{\overline{\alpha}_t*\sigma_0^2+1-\overline{\alpha}_t}+\sqrt{\overline{\alpha}_t}*\mu_0
\end{equation}
where $Mean(x)$ denotes the mean of $x$, and $Var(x)$ denotes the variance of $x$. Applying the normalized data for subsequent processing ensures both adherence to the standard curve and stability of the processing. Similarly, the restoration of the processed results can be expressed using the following formula:
\begin{equation}
Data_{restore}=\frac{x_{processed}-Mean(x_{processed})}{Var(x_{processed})}*Var(Data_{noisy})+Mean(Data_{noisy})
\end{equation}

Thus, in this section, we have proposed a new method for normalization and restoration, enabling our proposed noise attenuation approach to adapt to various types of seismic data.

\section{Experiments}
\subsection{Methods for Comparison}

In our experiment, we selected four additional seismic data noise attenuation methods for comparison of experimental results.

The first method is f-x deconvolution (f-x deconv), which assumes that seismic data with a linear coaxial axis in the F-X domain can be represented by an autoregressive model for each frequency slice \cite[]{1}. In the field of seismic noise processing, f-x deconv is one of the most mature methods and is widely applied to practical exploration tasks, so we choose f-x deconv as a comparison method in our experiment. We split the data into multiple 40*40 patches and each with 80$\%$ overlaps with neighboring patches. And then, these patches are processed by f-x deconv and restored to the size of the original seismic data. The second method is Hankel sparse low-rank approximation (HSLR) \cite[]{38}, which has achieved success on various data.

The third method is a twice denoising autoencoder framework (TDAE), proposed by \cite{29}. Based on DAE \cite[]{39}, \cite{29}. added a data generator, in which local correlation (LC) is first developed to nonlinear LC to detect and extract the signal leakage.

The fourth method is based on DDPM. When using DDPM for seismic noise attenuation, the parameter $t$ related to the noise level cannot be calculated during the reverse process, making DDPM unable to be directly applied to seismic noise attenuation. \cite{32} proposed using PCA to estimate $t$, thereby attenuating noise through the reverse process. The experiments demonstrate that the method of estimating $t$ through PCA and then processing with DDPM can achieve better results than conventional computational methods and DL method, especially in strong noise scenarios.

\subsection{Experiment on Synthesis Data}

In this experiment, the synthetic data we used is sourced from \cite{37}. We selected 1000 samples from this dataset to form the training set, and an additional 100 samples were chosen to compose the validation set. Similar to DDPM, the proposed diffusion model with the fast reverse process (FDM) also requires the estimation of parameter $t$. Therefore, in this paper, we employ the PCA method\cite[]{40} used by \cite{32} to estimate $t$. We synthesized seismic data with four different levels of noise ($t=50, t=100, t=150, t=200$) and statistically analyzed the average SNR and time consumption of the results obtained by the five methods. Among them, HSLR and f-x deconv were implemented through MATLAB R2019a, and the processing time was measured by recording the CPU runtime. Due to the MATLAB code for patch segmentation and Python for GPU computing simultaneously, we also measured the CPU runtime of TDAE for comparison. Since DDPM and FDM mainly utilize GPU for reverse computations, we recorded their GPU runtime for comparison. The average SNR and time consumption of the results obtained by five methods are shown in Figure ~\ref{fig:4}.

It can be observed that in cases of strong noise (SNR$<0$), both conventional computational methods and TDAE fail to achieve satisfactory results, while the SNR of the results obtained by FDM and DDPM is significantly higher than the other three methods. Even at lower levels of noise (SNR$>5$), although the performance gap among various methods is not significant, FDM and DDPM still outperform the other comparative methods. Compared to DDPM, due to the abandonment of random terms in the calculation formula during the inversion process, FDM consistently outperforms DDPM across different levels of noise in the experiments.

In comparing the time consumed by various methods to process the validation set, DDPM consumes a significant amount of time due to its step-by-step computations during the reverse process. Due to the patch segmentation and self-supervised training steps, TDAE also consumed a significant amount of time that is comparable to the time consumed by HSLR, which utilizes sparse transformation and low-rank computation. As the most widely used and mature, f-x deconv has the shortest processing time. Although our proposed FDM is slightly slower than f-x deconv, it outperforms the other three methods by several times, and the quality of the processed results is also the best.

Additionally, we extracted several data for visualization, as shown in Figure ~\ref{fig:5}. It is evident that on data with low noise levels, the FDM processing results in the highest SNR. At the same time, there are some artifacts present in the results of HSLR and f-x deconv, leading to lower SNR. While DDPM and TDAE can achieve results superior to the other two conventional computational methods, they still exhibit certain limitations compared to FDM. However, when the noise level is very high, TDAE, HSLR, and f-x deconv all fail to produce satisfactory results. There are residual noise and artifacts that have not been completely attenuated in their results. Both FDM and DDPM can achieve good results, with FDM showing the best performance among them.

Meanwhile, we compared the residual maps of various methods, with the calculation formula being:
\begin{equation}
residual\ map =processed\ data-clean\ data
\end{equation}

It is more evident from the residual maps that the results obtained by FDM are closest to the ground truth. Especially when the noise level is very high, the noise completely masks the waveform of the effective signal, leading to waveform reflections in the residual maps. In the results of f-x deconv, HSLR, and TDAE, a significant amount of artifacts or signal leakage is present. In such scenarios, FDM is still able to achieve more satisfactory results with fewer residuals.

Based on these experimental results, we can conclude that our proposed FDM method not only outperforms the comparative methods in terms of computational efficiency and processing performance but also can handle data with extremely high noise levels. To further substantiate our experimental conclusions, in the next section we will conduct experiments using actual seismic data for comparison.

\subsection{Experiment on Field Data}
\subsubsection{Transfer Learning}
Transfer learning is a commonly used technique in DL methods. By training on synthetic datasets and then transferring the pre-trained model to field data for validation, it effectively reduces the requirements for field data and labeled data \cite[]{41}. Due to the high cost of field exploration, which is always accompanied by a certain level of noise, acquiring a large amount of training data is extremely challenging. Therefore, in this study, we first trained the model on synthetic data and then transferred it to field data for validation.

\subsubsection{Pre-stack Seismic Gather}

The first data we selected is pre-stack seismic gather. Due to the lack of stacking, pre-stack seismic gather typically exhibits strong noise. The seismic gather has a sampling time interval of 2ms, with a single coverage of 61 seismic traces, totaling 30 coverages. We extracted a single coverage from the processed data for visualization, as shown in Figure ~\ref{fig:6}. As can be seen, all five methods have achieved good results. Due to more iterations and the presence of random terms in the iteration equation, artifacts appear in the result of DDPM, which can be trimmed by comparison with the original noisy data.

In addition, we selected two relatively typical areas for a more rigorous analysis. The comparison of the first area is shown in Figure ~\ref{fig:7}. In this area, the random noise is weaker compared to the effective seismic signal. The results of FDM, DDPM, and f-x deconv are similar, as they have eliminated most of the noise while preserving the effective signal intact. However, TDAE and HSLR show some residual noise in their results, making them less effective compared to the other three methods.

In the second area, the strength of the effective seismic signal is low, indicating strong interference from random noise, as shown in Figure ~\ref{fig:8}. It can be observed that the results of FDM and DDPM are similar, with FDM providing more detailed information. However, the results from TDAE and HSLR still retain significant noise, which is not conducive to subsequent data processing. Although f-x deconv has eliminated the majority of noise, it has also noticeably weakened the effective signal, leading to instances of missing effective signal. In addition to comparing the noise attenuation effects, we also compared the processing time consumed by each of the five methods when handling single coverage gather (FDM: 10.8633s; DDPM: 104.1716s; TDAE: 97.3618s; HSLR: 3.8806s; f-x deconv: 7.4804s). The calculation speed of the two conventional computational methods is faster than the other three methods, with our proposed FDM being the next fastest. Although the conventional computational methods are faster, their performance is slightly inferior compared to the other three methods. FDM has the best performance, and while its computation time is slightly slower than the conventional computational methods, it shows a 9 to 10-fold improvement compared to TDAE and DDPM. When processing more seismic data, this difference will further widen.

\subsubsection{USGS Central Alaska Dataset}

In our experiment, the data we used comes from USGS Central Alaska, which consists of 2D seismic data from 21 survey lines. We selected the 12th survey line, which contains 5582 traces and has a time sampling interval of 4ms and a recording duration of 6 seconds. Within the entire survey line, we selected a section with high noise levels, which includes seismic traces from 4001 to 5280, spanning a time range from 0.284s to 1.304s. We also applied the five methods individually to process this data, and the results are shown in Figure ~\ref{fig:9}. It is noticeable that FDM, DDPM, and TDAE have identified a significant amount of noise, while the result of f-x deconv attenuated the effective signal. In contrast to the other four methods, HSLR exhibits evident signal leakage in its results. Besides, due to the large horizontal scale of the data, to more intuitively compare the processing effects of the five methods, we selected a region with the strongest noise for display, as shown in Figure ~\ref{fig:10}.

The results of TDAE, HSLR, and f-x deconv show a significant amount of signal loss, while the methods of DDPM and FDM can eliminate a large amount of noise while completely preserving the effective signal. Compared to DDPM, the results from FDM exhibit richer details. In particular, the results of the two conventional computational methods exhibit severe signal loss, which significantly impacts subsequent data processing work. To evaluate the signal leakage of the five methods, we used local similarity maps\cite{42} for assessment, which is shown in Figure ~\ref{fig:11}.

The stronger the local similarity, the more significant the signal leakage. The results from HSLR exhibit extremely strong signal leakage, followed by f-x deconv. These leaked signals will significantly impact the subsequent works. Our proposed FDM, compared to DDPM and TDAE, not only identifies a large amount of noise but also more completely preserves the effective signals. Moreover, we conducted a statistical analysis of the time consumed by the five methods in processing this data (FDM: 7.4955s; DDPM: 47.8520s; TDAE: 117.4543s; HSLR: 279.4942s; f-x deconv: 9.4445s). Due to the enormous number of seismic traces, the time consumed by HSLR also increases sharply. The most effective FDM method has the shortest processing time, followed by f-x deconv. Compared to TDAE and DDPM, FDM shows an improvement of approximately 6-17 times.

\section{Discussion}

In this section, we will derive the SNR curve corresponding to different $t$ scenarios and we believe that can be directly obtained by statistical analysis of the training set.

Firstly, the expression for SNR is as follows:
\begin{equation}
SNR=10*\log_{10}{\frac{\sum_{i=1}^{N} x_0^2}{\sum_{i=1}^{N} (x_t-x_0)^2}}
\end{equation}

According to Equation 7 and 28, we further simplify the SNR formula:
\begin{equation}
\sum_{i=1}^{N}x_0*\hat{z}_t=\mu_0*\sum_{i=1}^{N}\hat{z}_t+\sigma_0*\sum_{i=1}^{N}(e_0*\hat{z}_t)
\end{equation}
as $e_0$ and $\hat{z}_t$ are independent standard normal distributions, the mean of the new distribution resulting from their multiplication is 0. Therefore, Equation 36 equals 0. And then, since the square of a standard normal distribution corresponds to a chi-square distribution, and the mean of a chi-square distribution is equal to its degrees of freedom, therefore:
\begin{equation}
(1-\overline{\alpha}_t)*\sum_{i=1}^{N}\hat{z}_t^2=(1-\overline{\alpha}_t)*N
\end{equation}
\begin{equation}
(1-\sqrt{\overline{\alpha}_t})^2*\sum_{i=1}^{N}x_0^2=(1-\sqrt{\overline{\alpha}_t})^2*(\mu_0^2+\sigma_0^2)*N
\end{equation}

In conclusion, the expression for SNR is:
\begin{equation}
SNR=10*\log_{10}{\frac{\mu_0^2+\sigma_0^2}{(1-\sqrt{\overline{\alpha}_t})^2*(\mu_0^2+\sigma_0^2)+(1-\overline{\alpha}_t)}}
\end{equation}

This curve is only related to the training dataset and represents the theoretical SNR improvement curve. We experimented by adding forward noise to a synthetic dataset, and the results generally align with the aforementioned SNR curve. However, due to the limitations of the DL model's accuracy, there is some error between the predicted $\hat{z}_t$ and the actual situation at each $t$, which hinders the smooth increase of SNR along this curve in our experiments. Especially when the noise level is too high, the results obtained in experiments often deviate significantly from the theoretical SNR. We believe that this issue can be improved through research in two aspects.

The first aspect is to improve the DL model to enhance its prediction accuracy. However, it is challenging to ensure both improved prediction accuracy and model efficiency simultaneously.

The second aspect is to impose constraints on the DL model or the reverse process. In our experiments, we found that when the noise level is extremely high, the restoration of the signal becomes very difficult due to the effective signal being almost destroyed. In this case, whether it is DDPM or FDM, the processed results always contain some artifacts. Although these artifacts may appear to be valid seismic signals, they differ significantly from the target signal, which reduces the SNR of the processed results. We suspect that the errors generated by the DL model during the reverse process have caused a deviation in the overall direction of the reverse process. Therefore, we believe that constraining the reverse process or DL model to follow the SNR curve will be a very worthwhile direction for further research.

\section{Conclusion}
In this paper, we have proposed a novel fast diffusion model and normalization method for noise attenuation in seismic data. We propose using an improved reverse process of DDPM, achieved through enhancing the Bayesian equation to implement skip-step computation. By extracting a subchain from the Markov chain of DDPM for the reverse process, FDM effectively accelerates the computation speed. We conducted experiments on synthetic datasets and ultimately determined the optimal length of the subchain. Additionally, we have proposed an improved normalization and restoration method to reduce the difference between the training set and actual noisy data, enabling FDM to handle various types of seismic data. Whether on synthetic datasets or field datasets, when facing strong noise, FDM not only outperforms the comparison methods in terms of processing effectiveness but also experiences a several-fold improvement in processing time. This has a very positive impact on improving the precision and efficiency of seismic exploration.

\section{ACKNOWLEDGMENTS}

We thank dGB Earth Sciences for making the data available as an OpendTect project via their TerraNubis portal \url{terranubis.com}. 

\newpage

\bibliographystyle{seg}  
\bibliography{example}

\renewcommand{\figdir}{figure} 
\newpage
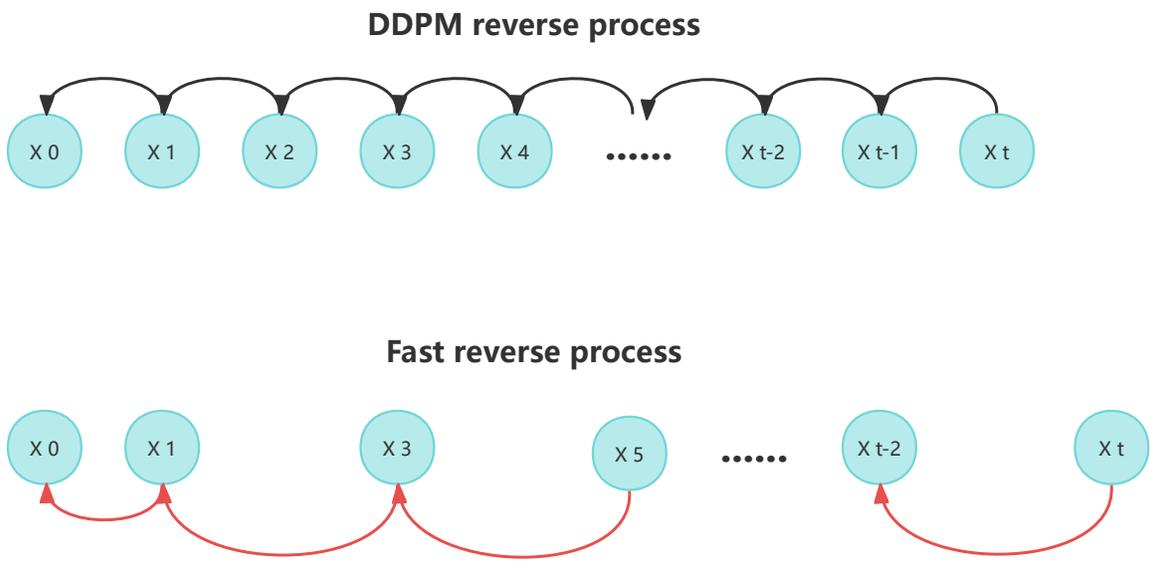
\plot{1}{width=1\textwidth}{The DDPM reverse process and the fast reverse process.}
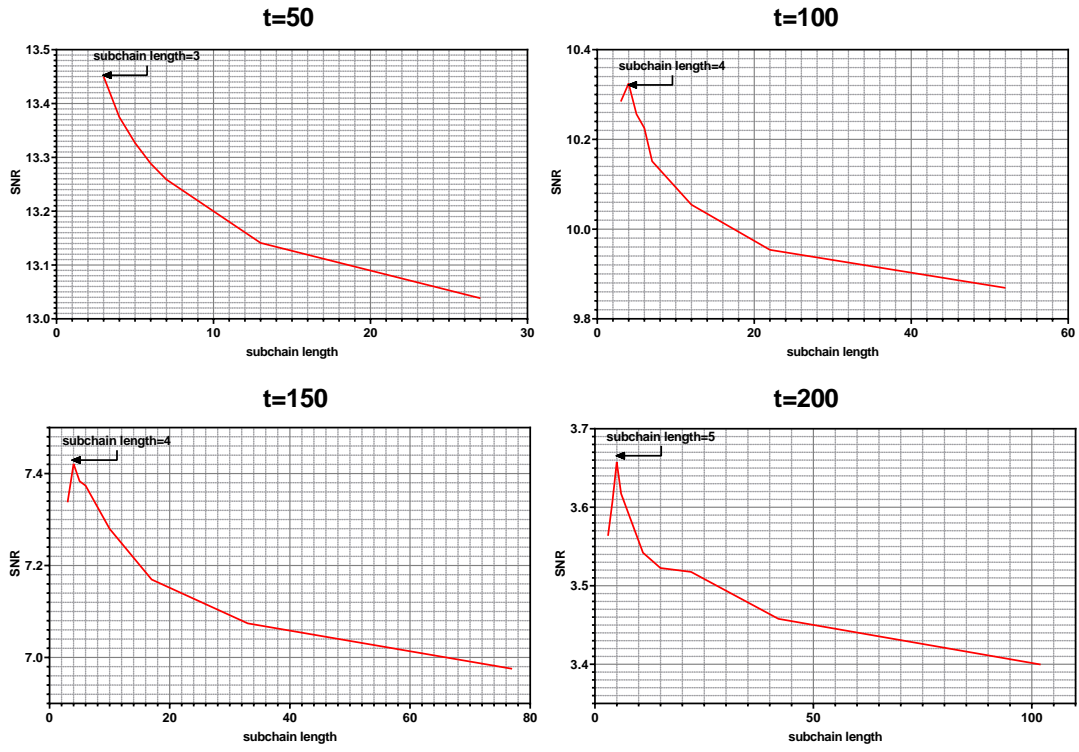
\plot{2}{width=1\textwidth}{The relationship between SNR and the length of the subchain.}
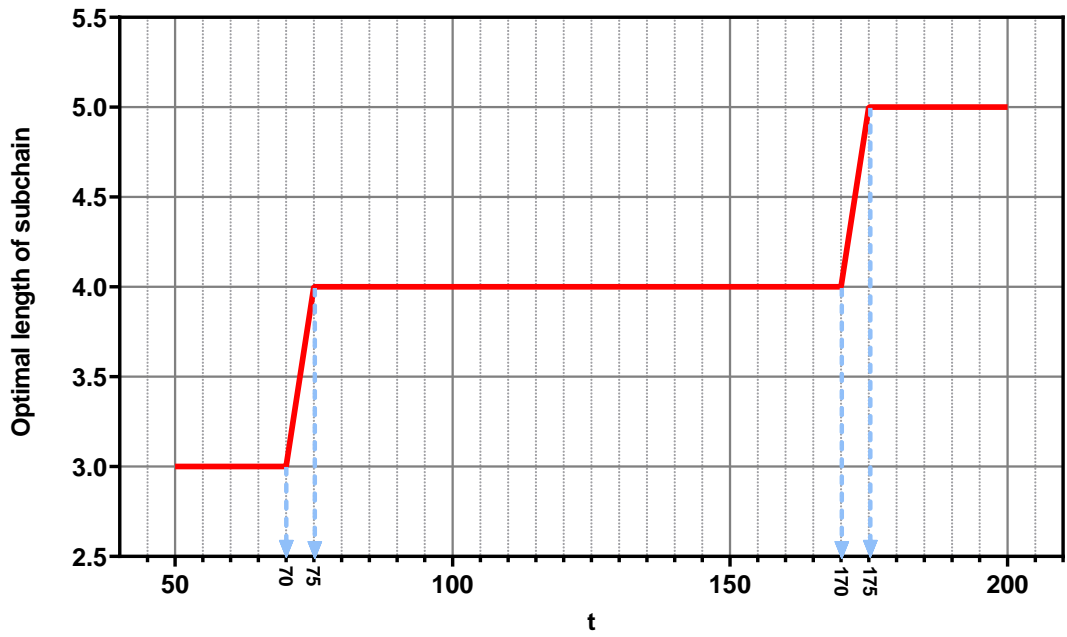
\plot{3}{width=1\textwidth}{The relationship between $t$ and the optimal length of the subchain.}
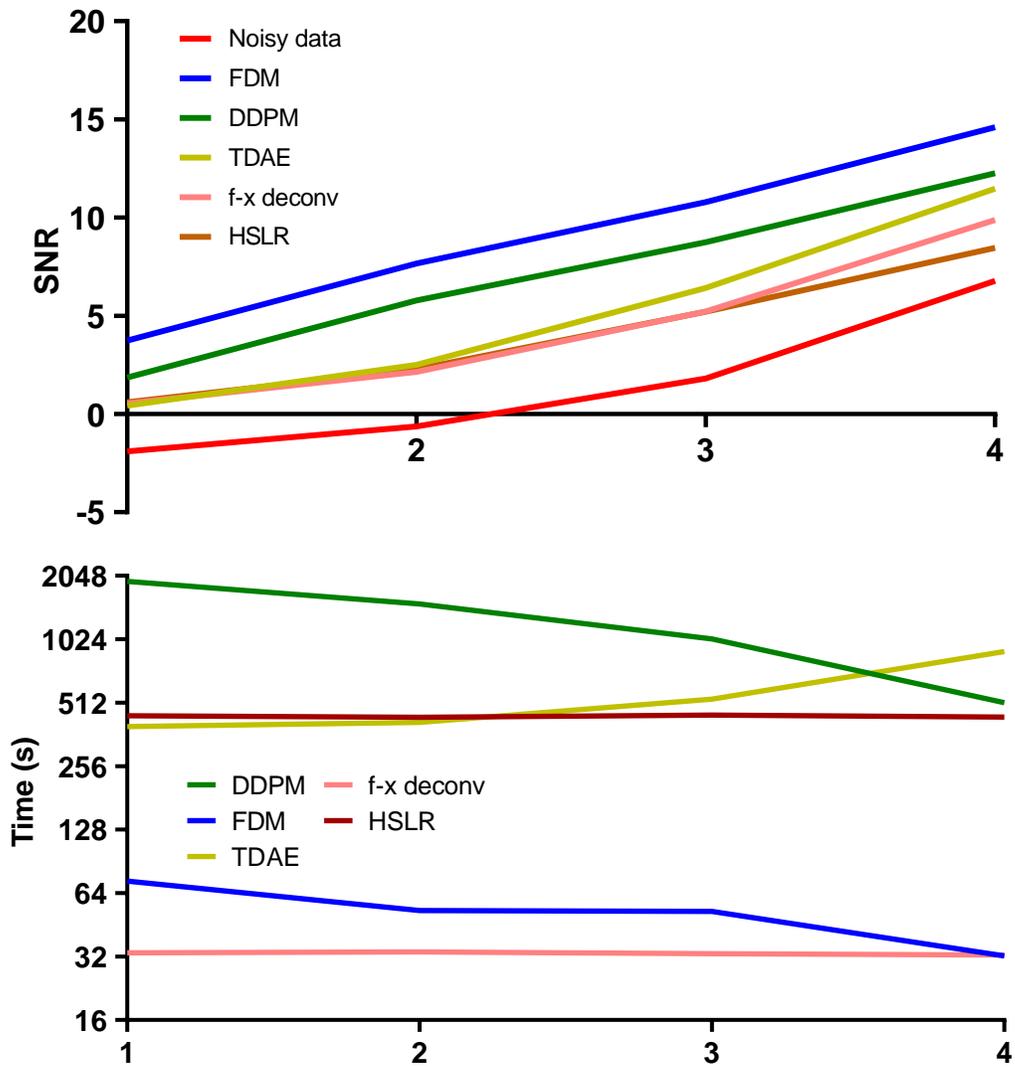
\plot{4}{width=1\textwidth}{The average SNR and time consumption of five methods on the validation set.}
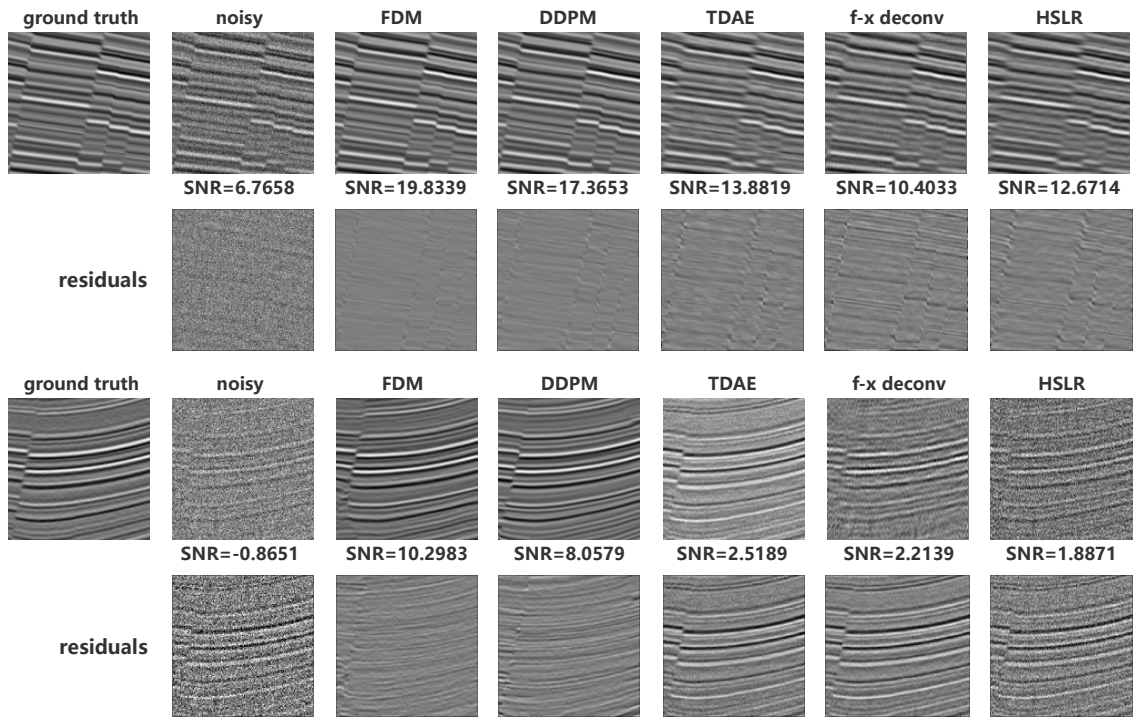
\plot{5}{width=1\textwidth}{Some examples of the results.}
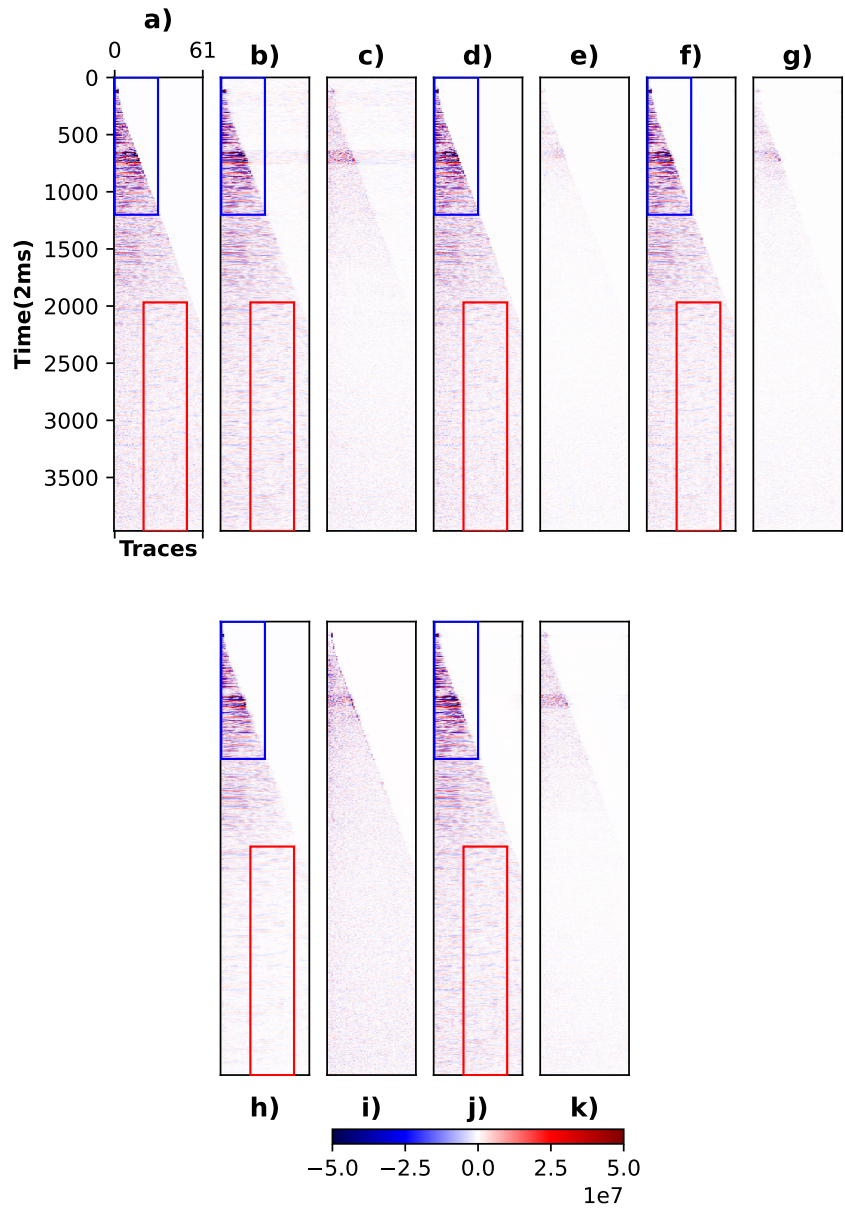
\plot{6}{width=0.75\textwidth}{Processed pre-stack seismic gather. a) original noisy gather; b) and c) represent the result of DDPM and its residual map; d) and e) represent the result of HSLR and its residual map; f) and g) represent the result of TDAE and its residual map; h) and i) represent the result of f-x deconv and its residual map; j) and k) represent the result of FDM and its residual map.}
\plot{7}{width=0.8\textwidth}{Processed pre-stack seismic gather. a) original noisy gather; b) the result of DDPM; c) the result of HSLR; d) the result of TDAE; e) the result of f-x deconv; f) the result of FDM.}
\plot{8}{width=0.8\textwidth}{Processed pre-stack seismic gather. a) original noisy gather; b) the result of DDPM; c) the result of HSLR; d) the result of TDAE; e) the result of f-x deconv; f) the result of FDM.}
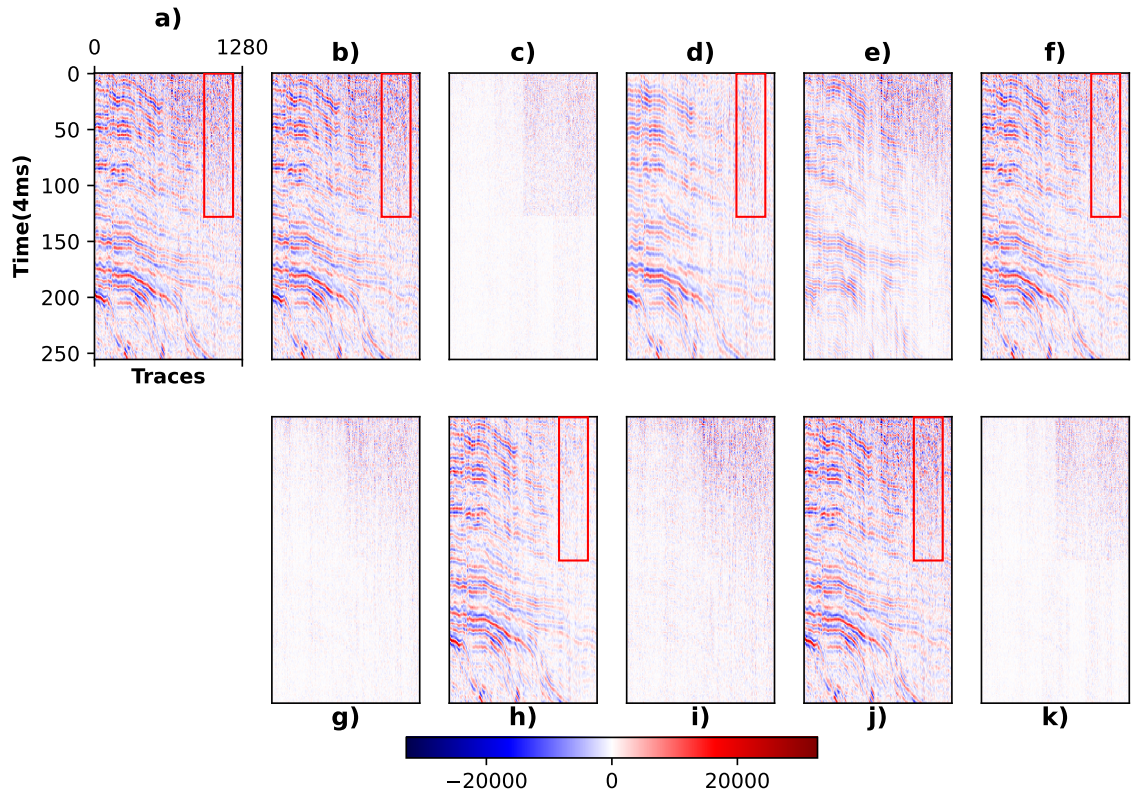
\plot{9}{width=1\textwidth}{Processed USGS Central Alaska Dataset. a) original noisy data; b) and c) represent the result of DDPM and its residual map; d) and e) represent the result of HSLR and its residual map; f) and g) represent the result of TDAE and its residual map; h) and i) represent the result of f-x deconv and its residual map; j) and k) represent the result of FDM and its residual map.}
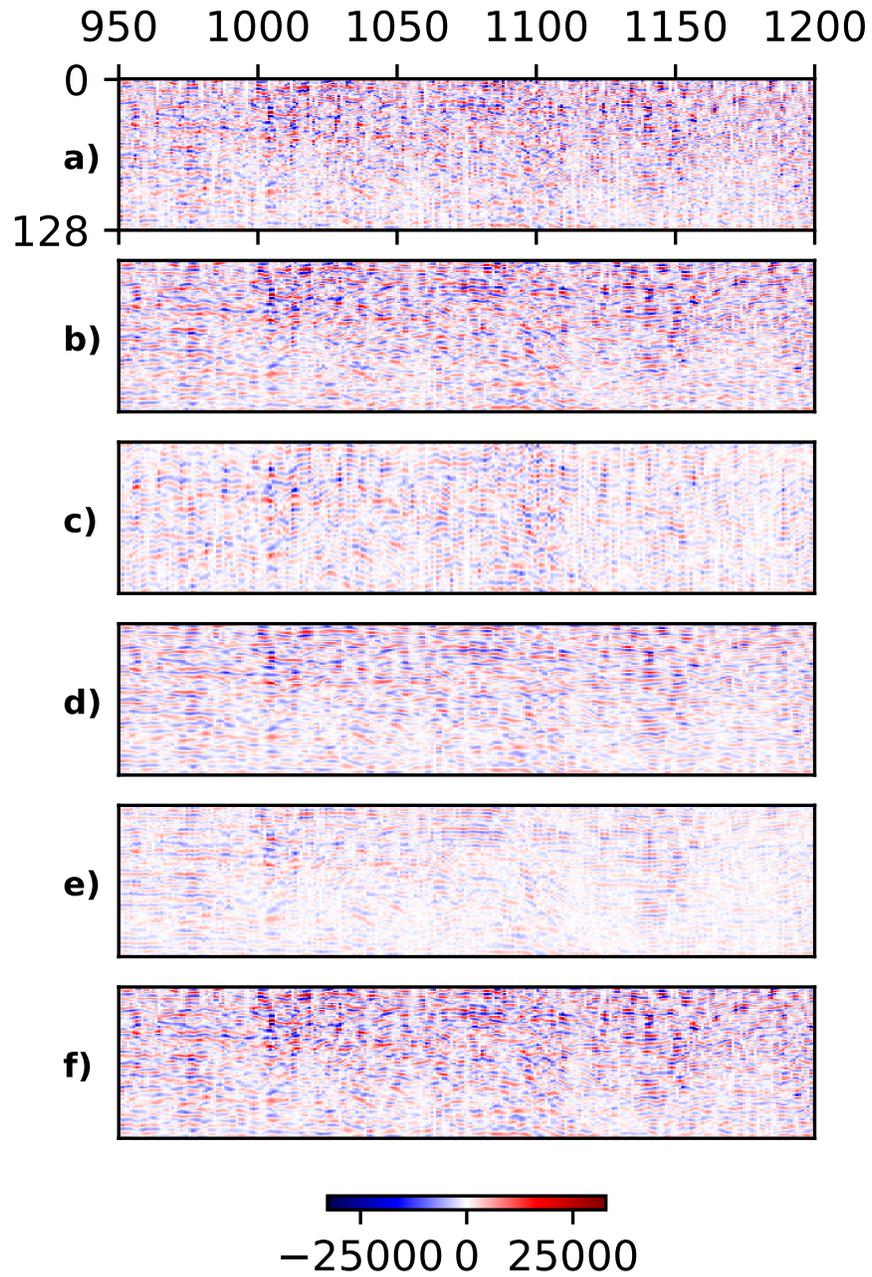
\plot{10}{width=0.8\textwidth}{Processed pre-stack seismic gather. a) original noisy data; b) the result of DDPM; c) the result of HSLR; d) the result of TDAE; e) the result of f-x deconv; f) the result of FDM.}
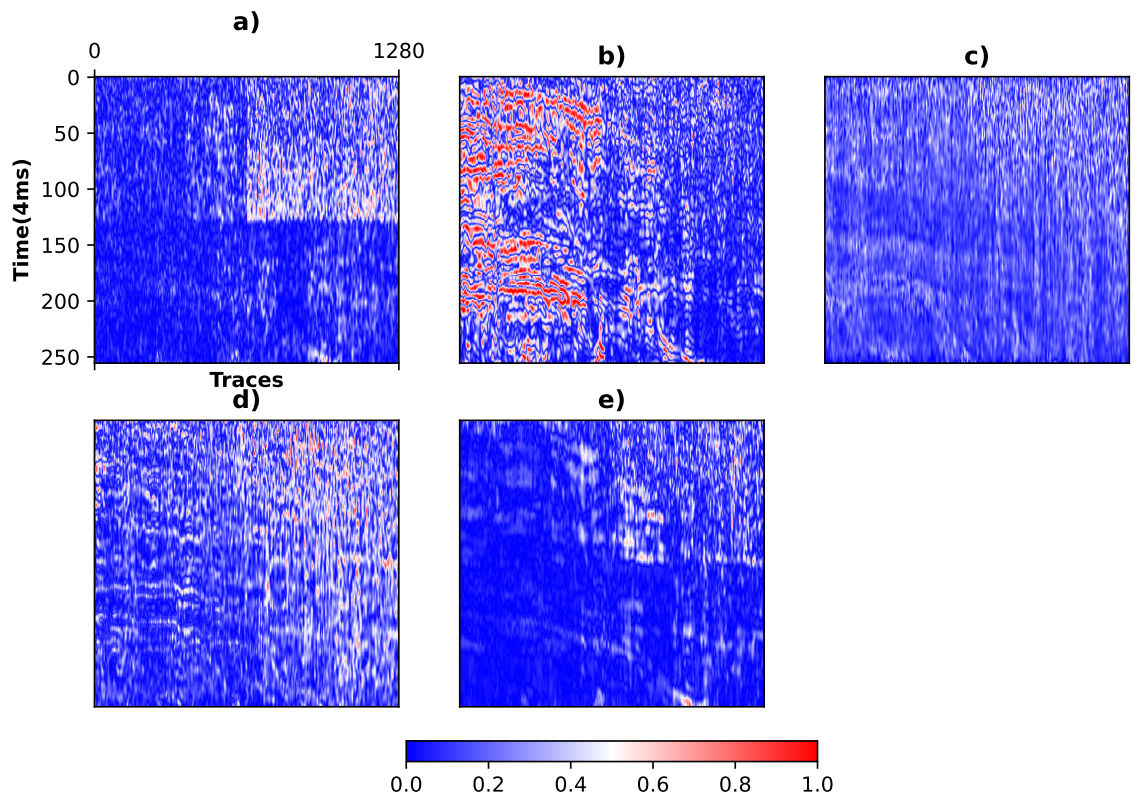
\plot{11}{width=1\textwidth}{Local similarity maps. a) the local similarity map of DDPM; b) the local similarity map of HSLR; c) the local similarity map of TDAE; d) the local similarity map of f-x deconv; e) the local similarity map of FDM.}

\end{document}